\input{aipcheck}

\documentclass[
    ,final            % use final for the camera ready runs
  ]
  {aipproc}

\layoutstyle{6x9}

\begin{document}

\title{General relativistic compact stars \\ with exotic matter}


\author{N. Yasutake}{
  address={Division of Theoretical Astronomy, National Astronomical Observatory of Japan, 2-21-1 Osawa, Mitaka, Tokyo 181-8588, Japan}
}

\author{T. Maruyama}{
  address={Advanced Science Research Center, Japan Atomic Energy Agency, Tokai, Ibaraki 319-1195, Japan}
}

\author{T. Tatsumi}{
  address={Department of Physics, Kyoto University, Kyoto 606-8502, Japan}
%  ,altaddress={<author1 address>} % additional visiting address
}

\author{K. Kiuchi}{
  address={Department of Physics, Waseda University, 3-4-1 Okubo, Shinjuku-ku, Tokyo 169-8555, Japan}
}

\author{K. Kotake}{
  address={Division of Theoretical Astronomy, National Astronomical Observatory of Japan, 2-21-1 Osawa, Mitaka, Tokyo 181-8588, Japan}
}

\begin{abstract}
We study structures of general relativistic compact stars with exotic matter. Our study is based on axisymetric and stationary formalism including purely toroidal magnetic field. We also study the finite size effects of quark-hadron mixed phase on structures of magnetars. For hybrid stars, we find a characteristic distribution of magnetic field, which has a discontinuity originated in the quark-hadron mixed phase. These distributions of magnetic field will change astrophysical phenomena, such as cooling processes.
\end{abstract}

\maketitle

%%%%%%%%%%%%%%%%%%%%%%%%%%%%%%%%%%%%%%%%%%%%
%% MAINMATTER
%%%%%%%%%%%%%%%%%%%%%%%%%%%%%%%%%%%%%%%%%%%%

\section{Introduction}
It was presented that quark matter may exist in compact stars~\cite{itoh70, bodmer71, witten84}. So far, there has been extensive work devoted to studying the effects of quark matter on astrophysical phenomena; the gravitational wave radiation~\cite{lin06, yasutake07, abdikamalov08},  cooling processes~\cite{page00,blaschke00,blaschke01,grigorian05}, neutrino emissions~\cite{nakazato08, sagert08}, and rotational frequencies~\cite{burgio03}, the maximum energy release by conversions from neutron stars to quark/hybrid stars~\cite{yasutake05, zdunik07}, etc.. However, uncertainties of equation of states~(EOSs) have been still left.

For such studies on compact stars, general relativistic effects are
fundamentally important, since baryon density is comparable to pressure,
$\rho_0 c^2 \sim P$, hence the gravitational forth is strong. Moreover, strong
magnetic field may change hydrostatic equilibriums of compact
stars. 

In this paper, we study the structures of general relativistic hybrid stars
with purely toroidal magnetic field. To calculate magnetized compact
stars, we adopt the scheme based on
axisymetric and stationary formalism including purely toroidal magnetic
field\cite{kiuchi08}. A star with pure toroidal magnetic field is unstable,
however it becomes another stable star in which toroidal magnetic field is
dominant in the dynamical simulation~\cite{kiuchi08b}. Moreover, Heger et al. have suggest that the toroidal magnetic field may dominate $10^5$ times lager than the poloidal magnetic field at the last stage of the main sequence~\cite{heger05}. 

The organization of this paper is as follows. Adopted EOSs are briefly discussed in Sec.~2 with  equilibriums of magnetized rotating compact
stars. In
Sec.~3, we show our nuemirical results. In Sec.~4, we discuss the consequence
of our calculations. 

%**********************
\section{Input physics}
%**********************
%*****************************
\subsection{Equation of state}
%*****************************
The hardness of EOSs is an important ingredient for determining the equilibrium configurations as mentioned in Sec.~1. Though proto-neutron stars left after supernova explosions are very hot, $T \sim 50$ MeV, they cool down to cold neutron stars ($T \sim 1 $MeV) in some tens of seconds~\cite{burrows86}. Therefore we assume that the temperature for hydrostatic compact stars is zero, since EOSs at such low temperature ($T \sim 1 $MeV) show almost same stiffness as at zero-temperature. It means the temperature of compact stars is always much smaller than the typical chemical potentials. Moreover, the loss of neutrinos means that the chemical potentials of the neutrinos may be set to zero. Thus, we impose the barotropic condition of the EOSs ($P=P(\epsilon)$) by assuming zero-temperature, zero-neutrino fraction, and beta-equilibrium. 

Our theoretical framework for the hadronic phase of matter is the
nonrelativistic Brueckner-Hartree-Fock approach including hyperons such as
$\Sigma^-$ and $\Lambda$~\cite{baldo98}.

For quark phase, we adopt the MIT bag model for quark phase of
u,~d~,s-quarks, because this is the first step for the study on strucures of hybrid stars with rotation and magnetic field in the general relativistic formulations, though it is a toy model. We use massless $u$ and $d$ quarks, and s-quark with a current mass of $m_s=180$ MeV. We set
that the bag constants, $B$, is 
100~MeV fm$^{-3}$ in this paper. Values of $B>150$ MeV~fm$^{-3}$ can also be
excluded within our model, because we do not obtain any more a phase
transition in beta-stable matter in combination with our hadronic EOS. 

For mixed phase, we must take into account the Gibbs condition, which require
the pressure balance and the equality of the chemical potentials between the
two phases besides the thermal equilibrium~\cite{heiselberg93}. We use the
Thomas-Fermi-approximation for the density plofiles of hadrons and quarks.
In each cell, we must calculate the balance of the colomb interaction and the
surface tension. However, there are a wide range of uncertainties about the
surface tension, $\sigma \sim 10-100$ MeV~fm$^{-2}$, which are suggested by some
theoretical estimates based on the MIT bag model for strangelets~\cite{farhi84} and lattice
guage simulations at finite temperature~\cite{huang90,kajantie91}. 
 
As for this uncertainity, Maruyama et al. have showed that the surface tensions of $\sigma > 40$ MeV~fm$^{-2}$ do not
change the hardness of EOSs~\cite{maruyama07}. Then, we adapt two values of surface
tension as $\sigma = 10$ MeV~fm$^{-2}$ for the minimum case, and $\sigma = 40$ MeV~fm$^{-2}$ for the maximum case.

At the maximum densities higher than two times of saturation density, muons may appear\cite{wiringa88, akmal98}. However, we neglect it, since the muon contribution to pressure at the higher density has been pointed to be very small~\cite{douchin01}.  

\begin{figure}
\includegraphics[width=58mm]{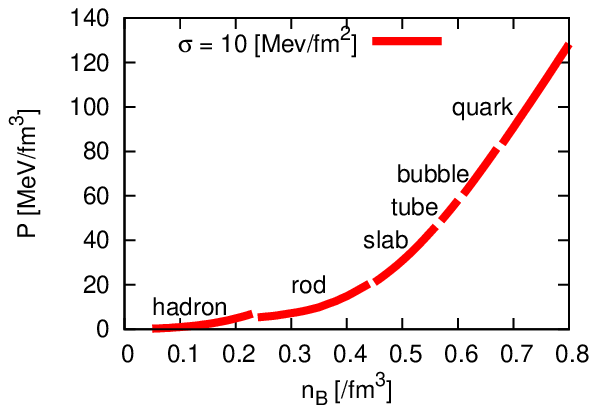}
\includegraphics[width=58mm]{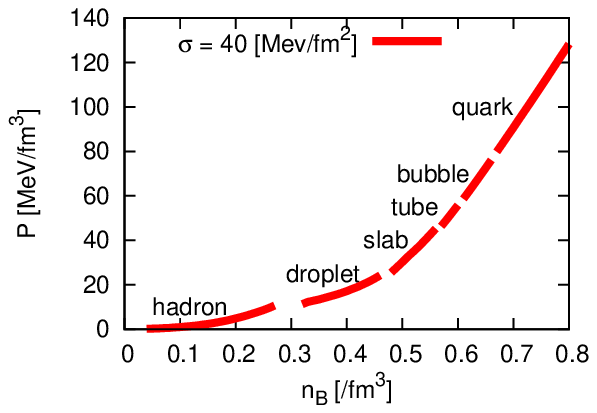}
\includegraphics[width=58mm]{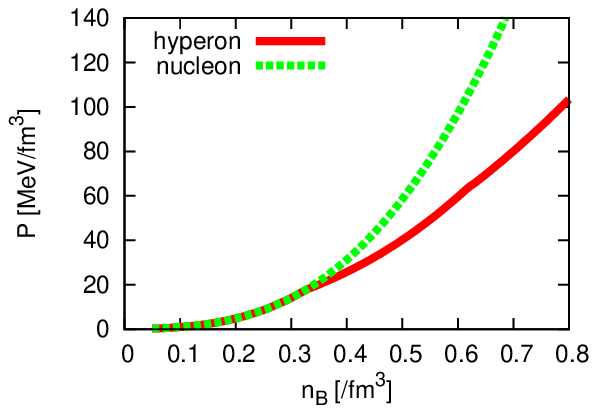}
\caption{\label{fig:eos_mix} Baryon number density versus pressures for each matter. The left~(center) panel is for an EOS with quark-hadron mixed phase under $B$=100 MeV fm$^{-3}$ and $\sigma$ = 10~(40) MeV~fm$^{-2}$. The right panel is for hadronic EOSs with~(without) hyperons shown as a solid~(dashed) line. Here, we assume neutrino free and zero temperature for all cases; $Y_{\nu_e}=0$, $T=0$. }
\end{figure}

Finally, we show our EOSs on Fig.~\ref{fig:eos_mix}. The left two panels show the pressure versus the baryon density for the quark-hadron mixed phase. For weak surface tension~($\sigma = 10$ MeV~fm$^{-2}$), the droplet structure does not appear, wheres, for strong surface tension~($\sigma = 40$ MeV~fm$^{-2}$), the rod structure does not appear. The number of hyperons are supressed because of the appearnce of quarks for both surface tensions. The right panel is for EOSs of nucleis with/without hyperons.  

%*************************************
\subsection{Equilibrium of magnetised rotating compact stars}
%*************************************
Master equations for the rotating relativistic stars containing purely toroidal magnetic fields are based on the assumptions summarized as follows~\cite{kiuchi08}; (1) Equilibrium models are stationary and axisymmetric. (2) The matter source is approximated by a perfect fluid with infinite conductivity. (3) There is no meridional flow of the matter. (4) The equation of state for the matter is barotropic. (5) Magnetic axis and rotation axis are aligned. This barotropic condition can be maintained for our EOSs. 

%***************************
\section{Numerical Results}\label{sec:result}
\begin{figure}
\includegraphics[width=13cm]{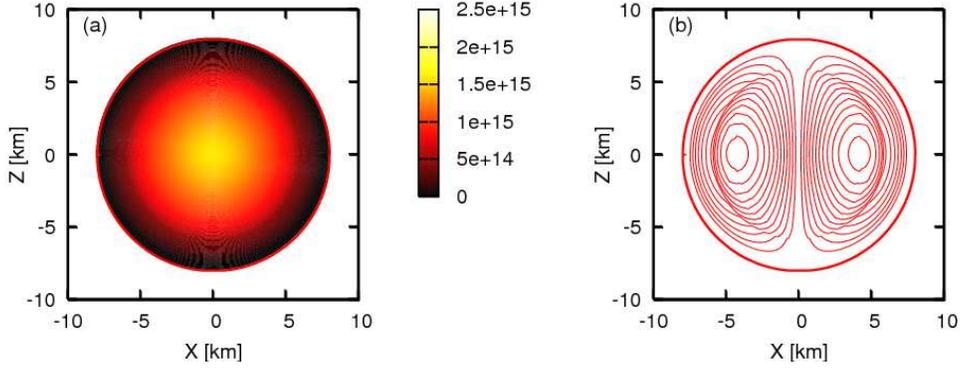} 
\caption{\label{fig:dis_mix} Distribution of (a): rest mass density [g/cm$^3$] and (b): magnetic field [G] with the quark-hadron phase transition($\sigma=40$ MeV/fm$^{-2}$).}
\end{figure}

\begin{figure}
\includegraphics[width=13cm]{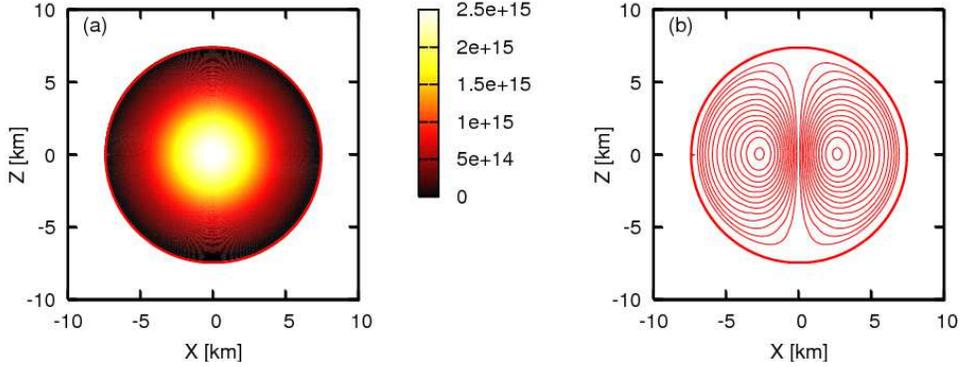} 
\caption{\label{fig:dis_nuc} Same as FIG.~\ref{fig:dis_mix} but without the
phase transition.}
\end{figure}

%***************************
Now, we show the configurations of strongly magnetized compact stars
with/without quark-hadron mixed phase. In this paper, we consider the
no-rotating static configurations for the following two reasons. (1) Since the magnetars
and the high field neutron stars observed so far are all slow rotators, the
static models could well approximate to such stars. (2) In the static models,
one can see purely magnetic effects on the equilibrium properties because
there is no centrifugal force and all the stellar deformation is attributed to
the magnetic stress. In Fig.~\ref{fig:dis_mix} and Fig.~\ref{fig:dis_nuc}, we
show distributions of the baryon density and the magnetic field in the
meridional planes for the two static equilibrium stars characterized by (1)
$\rho_{0,c}=1.58\times 10^{15}[{\rm g/cm^3}]$, $B_{max}=6.2 \times 10^{17}$~G, $M=1.31M_\odot$, $R_{\rm
cir}=9.97{\rm km}$, $H/|W|=2.11 \times 10^{-3}$ with the quark-hadron phase transition~($\sigma
=40$MeV/fm$^2$) [Fig.~\ref{fig:dis_mix}] and by (2)
$\rho_{0,c}=2.56\times 10^{15}[{\rm g/cm^3}]$, $B_{max}=7.1 \times 10^{17}$~G, $M=1.31M_\odot$, $R_{\rm
cir}=9.42{\rm km}$, $H/|W|=1.64 \times 10^{-3}$ of nucleon EOS with hyperons
[Fig.~\ref{fig:dis_nuc}]. Each physical value is as follows; the baryon mass
density $\rho_{0,c}$, the gravitational mass $M$, the circumferential radius
at the equator $R_{cir}$, the maximum strength of the magnetic fields
$B_{max}$, the ratio of the magnetic energy to the gravitational energy
$H/|W|$. These two models have a same magnetic flux of $5.00\times10^{29}{\rm G~cm^2}$ and a same baryon mass of $1.45 M_\odot$. 

Clearly, the distributions of magnetic field are different. The toroidal magnetic field lines behave like a rubber belt that is wrapped around the waist of the stars with hyperon EOS (see Fig.~\ref{fig:dis_nuc}). However, for a hybrid star, the distribution of magnetic field has discontinuity for the equatorial direction~(see Fig.~\ref{fig:dis_mix}). We can understand this easily; The magnetic field is frozen in the matter, hence the distribution of magnetic field depend on the distribution of the density. Hybrid stars have discontinuities of density profiles because of the phase transition; e.g. in FIG.~\ref{fig:dis_mix}, the 6 km from the core of a hybrid star is the mixed phase matter, and the baryon density in this density region has discontinuity, which raise the discontinuity of magnetic field as shown in the right panel of Fig.~2.

%**************************************************
\section{Summary and Discussion}\label{sec:summary}
%**************************************************
In this study, we have investigated the effects of quark-hadron mixed phase on the magnetized rotating stars of the general relativistic equilibrium configuration. As a result, we find that the distribution of magnetic field for hybrid stars has a discontinuity in the quark-hadron mixed phase. 

Let us move on effects of strong magnetic field on EOSs. 
It was pointed out that the strong magnetic field may change EOSs~\cite{broderick00}. In their study, they estimate the critical strength of magnetic field which may change EOSs, $\sim 10^{18}$~G. In our study, we calculate the structures of magnetized compact stars around $10^{18}$~G. Hence, such magnetic field may change EOSs. Especially for quark matter, it have founded that the energy gaps of magnetic Color-Flavor-Locked phase are oscillating functions of the magnetic field~\cite{noronha07, fukushima08}. Their effects are also important, and may change our results. 

As for the temperature and lepton fraction, we assume zero-temperature and
neutron free. We will check our result under finite temperature and neutrino
trapped cases. It will be useful for a study on structures of proto-compact
stars, or may  effect on cooling process and cooling curves of compact stars.

%%%%%%%%%%%%%%%%%%%%%%%%%%%%%%%%%%%%%%%%%%%%%%%%
%% BACKMATTER
%%%%%%%%%%%%%%%%%%%%%%%%%%%%%%%%%%%%%%%%%%%%%%%%

\begin{theacknowledgments}
We thank Y.~Eriguchi, S.~Yoshida for informative discussions.
\end{theacknowledgments}

%%%%%%%%%%%%%%%%%%%%%%%%%%%%%%%%%%%%%%%%%%%%%%%%
%% The bibliography can be prepared using the BibTeX program or
%% manually.
%%
%% The code below assumes that BibTeX is used.  If the bibliography is
%% produced without BibTeX comment out the following lines and see the
%% aipguide.pdf for further information.
%%
%% For your convenience a manually coded example is appended
%% after the \end{document}
%%%%%%%%%%%%%%%%%%%%%%%%%%%%%%%%%%%%%%%%%%%%%%%%

%%%%%%%%%%%%%%%%%%%%%%%%%%%%%%%%%%%%%%%%%%%%%%%%
%% You may have to change the BibTeX style below, depending on your
%% setup or preferences.
%%
%%
%% For The AIP proceedings layouts use either
%%%%%%%%%%%%%%%%%%%%%%%%%%%%%%%%%%%%%%%%%%%%

\bibliographystyle{aipproc}   % if natbib is available
%\bibliographystyle{aipprocl} % if natbib is missing

%%%%%%%%%%%%%%%%%%%%%%%%%%%%%%%%%%%%%%%%%%%
%% You probably want to use your own bibtex database here
%%%%%%%%%%%%%%%%%%%%%%%%%%%%%%%%%%%%%%%%%%%
\bibliography{sample}

\hyphenation{Post-Script Sprin-ger}
\begin{thebibliography}{31}
\expandafter\ifx\csname natexlab\endcsname\relax\def\natexlab#1{#1}\fi
\providecommand{\enquote}[1]{``#1''}
\expandafter\ifx\csname url\endcsname\relax
  \def\url#1{\texttt{#1}}\fi
\expandafter\ifx\csname urlprefix\endcsname\relax\def\urlprefix{URL }\fi
\providecommand{\eprint}[2][]{\url{#2}}

\bibitem[Itoh(1970)]{itoh70}
N.~Itoh, \emph{PTP} \textbf{44}, 291 (1970).

\bibitem[Bodmer(1971)]{bodmer71}
A.~R. Bodmer, \emph{PRD} \textbf{4}, 160 (1971).

\bibitem[Witten(1984)]{witten84}
E.~Witten, \emph{PDR} \textbf{30}, 272 (1984).

\bibitem[Lin et~al.(2006)]{lin06}
L.~M. Lin, K.~S. Cheng, M.~C. Chu, and W.~M. Suen, \emph{ApJ} \textbf{639}, 382
  (2006).

\bibitem[Yasutake et~al.(2007)]{yasutake07}
N.~Yasutake, K.~Kotake, M.~Hashimoto, and S.~Yamada, \emph{PDR} \textbf{75},
  084012 (2007).

\bibitem[Abdikamalov et~al.(2008)]{abdikamalov08}
E.~B. Abdikamalov, H.~Dimmelmeier, L.~Rezzolla, and J.~C. Miller,
  \emph{astro-ph/} \textbf{0806}, 1700 (2008).

\bibitem[Page et~al.(2000)]{page00}
D.~Page, M.~Prakash, J.~M. Lattimer, and A.~W. Steiner, \emph{PRL} \textbf{85},
  2048 (2000).

\bibitem[Blaschke et~al.(2000)]{blaschke00}
D.~Blaschke, T.~Klahn, and D.~N. Voskresensky, \emph{ApJ} \textbf{533}, 406
  (2000).

\bibitem[Blaschke et~al.(2001)]{blaschke01}
D.~Blaschke, H.~Grigorian, and D.~Voskresensky, \emph{A\&A} \textbf{368}, 561
  (2001).

\bibitem[Grigorian et~al.(2005)]{grigorian05}
H.~Grigorian, D.~Blaschke, and D.~Voskresensky, \emph{PRC} \textbf{71}, 045801
  (2005).

\bibitem[Nakazato et~al.(2008)]{nakazato08}
K.~Nakazato, K.~Sumiyoshi, and S.~Yamada, \emph{PRD} \textbf{77}, 103006
  (2008).

\bibitem[Sagert et~al.(2008)]{sagert08}
I.~Sagert, M.~Hempel, G.~Pagliara, J.~Schaffner-Bielich, T.~Fischer,
  A.~Mezzacappa, F.~K. Thielemann, and M.~Liebend$\ddot{o}$rfer,
  \emph{astro-ph/} \textbf{0809}, 4225 (2008).

\bibitem[Burgio et~al.(2003)]{burgio03}
G.~F. Burgio, H.~J. Schulze, and F.~Weber, \emph{A\&A} \textbf{408}, 675
  (2003).

\bibitem[Yasutake et~al.(2005)]{yasutake05}
N.~Yasutake, M.~Hashimoto, and Y.~Eriguchi, \emph{PTP} \textbf{113}, 953
  (2005).

\bibitem[Zdunik et~al.(2007)]{zdunik07}
J.~L. Zdunik, M.~Bejger, P.~Haensel, and E.~Gourgoulhon, \emph{A\&A}
  \textbf{465}, 533 (2007).

\bibitem[Kiuchi and Yoshida(2008)]{kiuchi08}
K.~Kiuchi, and S.~Yoshida, \emph{PRD} \textbf{78}, 044045 (2008).

\bibitem[Kiuchi et~al.(2008)]{kiuchi08b}
K.~Kiuchi, M.~Shibata, and S.~Yoshida, \emph{astro-ph/} \textbf{0805}, 2712
  (2008).

\bibitem[Heger et~al.(2005)]{heger05}
A.~Heger, S.~E. Woosley, and H.~C. Spruit, \emph{ApJ} \textbf{626}, 350 (2005).

\bibitem[Burrows and Lattimer(1986)]{burrows86}
A.~Burrows, and J.~M. Lattimer, \emph{ApJ} \textbf{307}, 178 (1986).

\bibitem[Baldo et~al.(1998)]{baldo98}
M.~Baldo, G.~F. Burgio, and H.-J. Schulze, \emph{PRC} \textbf{58}, 3688 (1998).

\bibitem[Heiselberg et~al.(1993)]{heiselberg93}
H.~Heiselberg, C.~J. Pethick, and E.~F. Staubo, \emph{PRL} \textbf{70}, 1355
  (1993).

\bibitem[Farhi and Jaffe(1984)]{farhi84}
E.~Farhi, and R.~L. Jaffe, \emph{PRD} \textbf{30}, 2379 (1984).

\bibitem[Huang et~al.(1990)]{huang90}
S.~Huang, J.~Potvion, C.~Rebbi, and S.~Sanielevici, \emph{PRD} \textbf{42},
  2864 (1990).

\bibitem[Kajantie et~al.(1991)]{kajantie91}
K.~Kajantie, L.~K$\ddot{a}$rk$\ddot{a}$inen, and K.~Rummukainen, \emph{Nucl.
  Phys. B} \textbf{357}, 693 (1991).

\bibitem[Maruyama et~al.(2007)]{maruyama07}
T.~Maruyama, S.~Chiba, H.-J. Schulze, and T.~Tatsumi, \emph{PRD} \textbf{76},
  123015 (2007).

\bibitem[Wiringa et~al.(1988)]{wiringa88}
R.~B. Wiringa, V.~Filks, and A.~Fabrocini, \emph{PRC} \textbf{38}, 1010 (1988).

\bibitem[Akmal et~al.(1998)]{akmal98}
A.~Akmal, V.~R. Pandharipande, and D.~G. Ravenhall, \emph{PRC} \textbf{58},
  1804 (1998).

\bibitem[Douchin and Haencel(2001)]{douchin01}
F.~Douchin, and P.~Haencel, \emph{A\&A} \textbf{380}, 151 (2001).

\bibitem[Broderick et~al.(2000)]{broderick00}
A.~Broderick, M.~Prakash, and J.~M. Lattimer, \emph{ApJ} \textbf{537}, 351
  (2000).

\bibitem[Noronha and Shovkovy(2007)]{noronha07}
J.~L. Noronha, and I.~A. Shovkovy, \emph{PRD} \textbf{76}, 105030 (2007).

\bibitem[Fukushima and Warringa(2008)]{fukushima08}
K.~Fukushima, and H.~J. Warringa, \emph{PRD} \textbf{78}, 039902 (2008).

\end{thebibliography}

%%%%%%%%%%%%%%%%%%%%%%%%%%%%%%%%%%%%%%%%%%%
%% Just a reminder that you may have to run bibtex
%% All of it up to \end{document} can be removed
%% if you don't like the warning.
%%%%%%%%%%%%%%%%%%%%%%%%%%%%%%%%%%%%%%%%%%%
\IfFileExists{\jobname.bbl}{}
 {\typeout{}
  \typeout{******************************************}
  \typeout{** Please run "bibtex \jobname" to optain}
  \typeout{** the bibliography and then re-run LaTeX}
  \typeout{** twice to fix the references!}
  \typeout{******************************************}
  \typeout{}
 }

\end{document}